\documentclass[twocolumn,pre,superscriptaddress]{revtex4}

\usepackage{graphicx}
\usepackage{amsmath,amssymb}
\usepackage{calrsfs}
\usepackage[loose]{subfigure}

\newcommand\defn{\textit}
\newcommand\mat{\mathbf}
\newcommand\e{\mathrm{e}}

\renewcommand\vec{\mathbf}

\newcommand\half{\tfrac12}
\newcommand{\Ord}{\mathrm{O}}

\newcommand{\tr}{\operatorname{Tr}}
\newcommand{\bm}{\boldsymbol}

\begin{document}

\title{First-principles multiway spectral partitioning of graphs}

\author{Maria A. Riolo}
\affiliation{Department of Mathematics, University of Michigan, Ann Arbor,
  MI 48109}
\affiliation{Center for the Study of Complex Systems, University of
  Michigan, Ann Arbor, MI 48109}
\author{M. E. J. Newman}
\affiliation{Center for the Study of Complex Systems, University of
  Michigan, Ann Arbor, MI 48109}
\affiliation{Department of Physics, University of Michigan, Ann Arbor, MI
  48109}

\begin{abstract}
  We consider the minimum-cut partitioning of a graph into more than two
  parts using spectral methods.  While there exist well-established
  spectral algorithms for this problem that give good results, they have
  traditionally not been well motivated.  Rather than being derived from
  first principles by minimizing graph cuts, they are typically presented
  without direct derivation and then proved after the fact to work.  In
  this paper, we take a contrasting approach in which we start with a
  matrix formulation of the minimum cut problem and then show, via a
  relaxed optimization, how it can be mapped onto a spectral embedding
  defined by the leading eigenvectors of the graph Laplacian.  The end
  result is an algorithm that is similar in spirit to, but different in
  detail from, previous spectral partitioning approaches.  In tests of the
  algorithm we find that it outperforms previous approaches on certain
  particularly difficult partitioning problems.
\end{abstract}


\maketitle

\section{Introduction}
Graph partitioning, the division of a graph or network into weakly
connected subgraphs, is a problem that arises in many areas, including
finite element analysis, electronic circuit design, parallel computing,
network data analysis and visualization, and
others~\cite{Elsner97,Fjallstrom98}.  In the most basic formulation of the
problem one is given an undirected, unweighted graph of $n$ vertices and
asked to divide the vertices into $k$ nonoverlapping groups of given sizes,
such that the number of edges running between groups---the so-called
\defn{cut size}---is minimized.  This is known to be a computationally hard
problem.  Even for the simplest case where $k=2$ it is NP-hard to find the
division with minimum cut size~\cite{GJ79}.  Good approximations to the
minimum can, however, be found using a variety of heuristic methods,
including local greedy algorithms, genetic algorithms, tabu search, and
multilevel algorithms.  One particularly elegant and effective approach,
which is the subject of this paper, is \defn{spectral
  partitioning}~\cite{vonLuxburg07}, which makes use of the spectral
properties of any of several matrix representations of the graph, most
commonly the graph Laplacian.

The first Laplacian spectral partitioning algorithms date back to the work
of Fiedler in the 1970s~\cite{Fiedler73,PSL90} and were aimed at solving
the graph bisection problem, i.e.,~the problem of partitioning a graph into
just two parts.  For this problem the underlying theory of the spectral
method is well understood and the algorithms work well.  One calculates the
eigenvector corresponding to the second lowest eigenvalue of the graph
Laplacian and then divides the graph according to the values of the vector
elements---the complete process is described in
Section~\ref{sec:bisection}.  More recently, attention has turned to the
general multiway partitioning problem with arbitrary~$k$, which is harder.
One elementary approach is to repeatedly bisect the graph into smaller and
smaller parts using Fiedler's method or one of its variants, but this can
give rise to poor solutions in some commonly occurring situations.
A~better approach, and the one in widest current use, is to construct the
$n\times(k-1)$ matrix whose columns are the eigenvectors for the second- to
$k$th-lowest eigenvalues of the Laplacian.  The rows of this matrix
define~$n$ vectors of $k-1$ elements each which are regarded as points in a
$(k-1)$-dimensional space.  One clusters these points into $k$ groups using
any of a variety of heuristics, the most common being the $k$-means method,
and the resulting clusters define the division of the graph.

This method works well in practice, giving good results on a wide range of
test graphs, and if one is concerned only with finding an algorithm that
works, one need look no further.  Formally, however, it does have some
drawbacks.  First, it is not a true generalization of the method for $k=2$.
If one were to apply this algorithm to a $k=2$ problem, one would end up
performing $k$-means on the second eigenvector of the graph Laplacian,
which is a different procedure from the standard $k=2$ algorithm.  Second,
the algorithm is not normally even derived directly from the minimum-cut
problem.  Instead, the algorithm is typically proposed without
justification, and justified after the fact by demonstrating that it
performs well on particular partitioning tasks.  This approach is perfectly
correct but somewhat unsatisfactory in that it does not give us much
understanding of why the calculation works.  For that, it would be better
to derive the algorithm from first principles.  The purpose of this paper
is to give such a derivation for multiway spectral partitioning.  Our main
goal in doing so is to gain an understanding of \emph{why} spectral
partitioning works, by contrast with the traditional presentation which
demonstrates only that it does work.  However, as we will see, the
algorithm that we derive in the process is different in significant ways
from previous spectral partitioning algorithms and in
Section~\ref{sec:results} we present results that indicate that our
algorithm can outperform more conventional approaches for certain classes
of graphs.

The previous literature on spectral partitioning is extensive---this has
been an active area of research, especially in the last few years.  There
is also a large literature on the related problem of spectral clustering of
high-dimensional data sets, which can be mapped onto a weighted
partitioning problem on a complete graph using an affinity matrix.  The
1995 paper of Alpert and Yao~\cite{AY95} provides an early example of an
explicit derivation of a general multiway partitioning algorithm.  Their
algorithm is substantially different from the most commonly used variants,
involving a vector partitioning step, and also contains one arbitrary
parameter which affects the performance of the algorithm but whose optimal
value is unknown.  The algorithms of Shi and Malik~\cite{SM00} and
Meil\u{a} and Shi~\cite{MS00} are good examples of the standard multiway
partitioning using $k$-means, although applied to the slightly different
problem of normalized-cut partitioning.  A number of subsequent papers have
analyzed these algorithms or variants of
them~\cite{NJW01,MX03,KKV04,LGT11}.  Summaries are given by von
Luxburg~\cite{vonLuxburg07}, Verma and Meil\u{a}~\cite{VM03b}, and Bach and
Jordan~\cite{BJ06}, although the discussions are in the language of data
clustering, not graph partitioning.  Perhaps the work that comes closest to
our own is that of Zhang and Jordan~\cite{ZJ08}, again on data clustering,
in which partitions are indexed using a set of $(k-1)$-dimensional ``margin
vectors,'' which are oriented using a Procrustes technique.  We also use
Procrustes analysis in one version of the method we describe, although
other details of our approach are different from the method of Zhang and
Jordan.

The outline of this paper is as follows.  In Section~\ref{sec:bisection} we
review the derivation of the standard spectral bisection algorithm and then
in Section~\ref{sec:kway} present in detail the generalization of that
derivation to the multiway partitioning problem, leading to an algorithm
for multiway partitioning of an arbitrary undirected graph into any number
of groups of specified sizes.  In Section~\ref{sec:results} we give example
applications of this algorithm to a number of test graphs, and demonstrate
that its performance is similar to, or in some cases slightly better than,
approaches based on $k$-means.  In Section~\ref{sec:conclusions} we give
our conclusions and discuss directions for future research.

\section{Spectral bisection}
\label{sec:bisection}
The term \defn{spectral bisection} refers to the special case in which we
partition a graph into exactly $k=2$ parts.  For this case there is a
well-established first-principles derivation of the standard partitioning
algorithm, which we review in this section.  Our goal in subsequent
sections will be to find a generalization to the case of arbitrary~$k$.

Suppose we are given an undirected, unweighted graph on $n$ vertices, which
we will assume to be connected (i.e.,~to have only one component), and we
wish to divide its vertices into two groups which, for the sake of
simplicity, we will take to be of equal size $\half n$ (with $n$ even).  We
define an index variable~$s_i$ for each vertex~$i=1\ldots n$ such that
$s_i=1$ if vertex~$i$ belongs to group~1 and $s_i=-1$ if $i$ belongs to
group~2.  We note that
\begin{equation}
\half(s_is_j+1) = \biggl\lbrace\begin{array}{ll}
                    1 & \quad\mbox{if $i$ and $j$ are in the same group,} \\
                    0 & \quad\mbox{otherwise.}
                  \end{array}
\label{eq:sdelta}
\end{equation}
Thus the number of edges within groups is given by $\frac12 \sum_{ij}
\half(s_is_j+1) A_{ij}$, where $A_{ij}$ is an element of the adjacency
matrix (having value~1 if there is an edge between $i$ and~$j$, and zero
otherwise) and the extra factor of~$\half$ compensates for double counting
of vertex pairs in the sum.  The total number of edges in the entire graph
is $\half\sum_{ij} A_{ij}$ and hence the number of edges between
groups---which is the cut size~$R$---is given by
\begin{align}
R &= \half\sum_{ij} A_{ij} - \tfrac14 \sum_{ij} (s_is_j+1) A_{ij}
     \nonumber\\
  &= \tfrac14 \sum_{ij} (1-s_is_j) A_{ij}
   = \tfrac14 \sum_i d_i - \tfrac14 \sum_{ij} s_is_j A_{ij},
\label{eq:r1}
\end{align}
where $d_i = \sum_j A_{ij}$ is the degree of vertex~$i$.  Noting that
$s_i^2=1$ for all~$i$, this equation can be rewritten as
\begin{equation}
R = \tfrac14 \sum_{ij} d_i \delta_{ij} s_i s_j
    - \tfrac14\sum_{ij} A_{ij} s_is_j
  = \tfrac14 \sum_{ij} L_{ij} s_i s_j,
\end{equation}
where $L_{ij} = d_i \delta_{ij} - A_{ij}$ is the $ij$th element of the
graph Laplacian matrix~$\mat{L}$.  Alternatively, we can write $R$ in
matrix notation as
\begin{equation}
R = \tfrac14 \vec{s}^T\mat{L}\vec{s},
\label{eq:cutsize0}
\end{equation}
where $\vec{s}$ is the $n$-component vector with elements~$s_i$.

\begin{figure}
\begin{center}
\includegraphics[width=4.5cm]{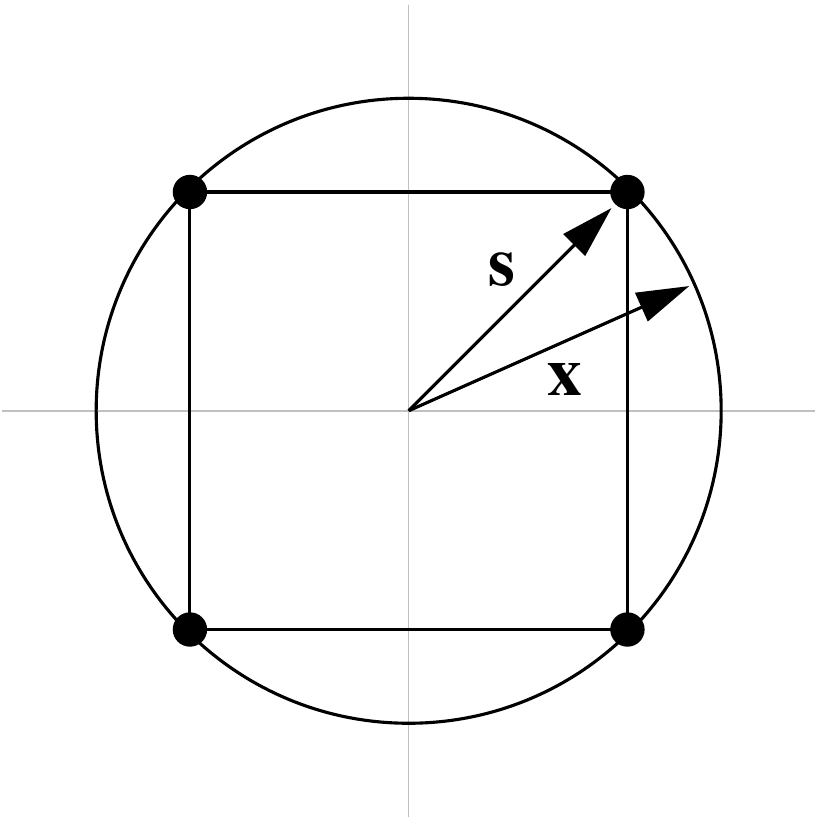}
\end{center}
\caption{The possible values of the vector~$\vec{s}$ lie at the corners of
  a hypercube in $n$-dimensional space, while the relaxed vector~$\vec{x}$
  can lie at any point on the circumscribing hypersphere.}
\label{fig:relax}
\end{figure}

Our goal, for a given graph and hence for given~$\mat{L}$, is to minimize
the cut size~$R$ over possible bisections of the graph, represented
by~$\vec{s}$, subject to the constraint that the two groups are the same
size, which is equivalent to saying that $\sum_i s_i = 0$ or
\begin{equation}
\vec{1}^T\vec{s} = 0,
\label{eq:ssum}
\end{equation}
where $\vec{1}$ is the uniform vector $(1,1,1,\ldots)$.  Unfortunately, as
mentioned in the introduction, this is a hard computational problem.  But
one can in many cases find good approximate solutions in polynomial time by
using a \defn{relaxation method}.  We generalize the discrete
variables~$s_i=\pm1$ to continuous real variables~$x_i$ and solve the
relaxed minimization with respect to the vector~$\vec{x}=(x_1,x_2,\ldots)$
of
\begin{equation}
R_x = \tfrac14 \vec{x}^T\mat{L}\vec{x},
\end{equation}
subject to the constraint
\begin{equation}
\vec{1}^T\vec{x} = 0,
\label{eq:sumx}
\end{equation}
which is the equivalent of Eq.~\eqref{eq:ssum}.  One must however also
apply an additional constraint to prevent $\vec{x}$ from becoming zero,
which is normally taken to have the form
\begin{equation}
\vec{x}^T\vec{x} = n.
\label{eq:sumxsq}
\end{equation}
Choices of $\vec{x}$ satisfying this second constraint include all allowed
values of the original unrelaxed vector~$\vec{s}$, since $\vec{s}^T\vec{s}
= \sum_i s_i^2 = \sum_i 1 = n$, but also include many other values in
addition.  Geometrically, one can think of $\vec{s}$ as defining a point in
an $n$-dimensional space, with the allowed values~$s_i=\pm1$ restricting
the point to fall at one of the corners of a hypercube.  The value
of~$\vec{x}$ falls on the circumscribing hypersphere, since
$\vec{x}^T\vec{x} = \sum_i x_i^2 = n$ implies that $\vec{x}$ has constant
length~$\sqrt{n}$.  The hypersphere coincides with the values of~$\vec{s}$
at the corners of the hypercube, but includes other values in between as
well---see Fig.~\ref{fig:relax}.

The relaxed problem is straightforward to solve by differentiation.  We
enforce the two conditions \eqref{eq:sumx} and \eqref{eq:sumxsq} with
Lagrange multipliers $\lambda$ and~$\mu$ so that
\begin{equation}
{\partial\over\partial x_k} \biggl[ \sum_{ij} L_{ij} x_i x_j
  - \lambda \sum_i x_i^2 - \mu \sum_i x_i \biggr] = 0.
\end{equation}
Performing the derivatives we find that $2\sum_j L_{kj} x_j - 2\lambda x_k
- \mu = 0$ or in matrix notation $\mat{L} \vec{x} = \lambda \vec{x} +
\half\mu \vec{1}$.  Multiplying on the left by~$\vec{1}$ we get
$\vec{1}^T\mat{L}\vec{x} = \lambda \vec{1}^T\vec{x} + \half n\mu$, and
employing Eq.~\eqref{eq:sumx} and noting that $\vec{1}$ is an eigenvalue of
$\mat{L}$ with eigenvalue zero, we find that $\mu=0$.  Thus we have
\begin{equation}
\mat{L}\vec{x} = \lambda\vec{x}.
\label{eq:solnx}
\end{equation}
In other words $\vec{x}$ is an eigenvector of the graph Laplacian
satisfying the two conditions \eqref{eq:sumx} and~\eqref{eq:sumxsq}.

Our solution is completed by noting that the cut size~$R_x$ within the
relaxed approximation, evaluated at the solution of~\eqref{eq:solnx}, is
\begin{equation}
R_x = \tfrac14 \vec{x}^T\mat{L}\vec{x} = \tfrac14 \lambda \vec{x}^T\vec{x}
    = {n\lambda\over 4}.
\label{eq:rx2}
\end{equation}
This is minimized by choosing $\lambda$ as small as possible, in other
words by choosing $\vec{x}$ to be the eigenvector corresponding to the
lowest possible eigenvalue.  The lowest eigenvalue of~$\mat{L}$ is always
zero, with corresponding eigenvector proportional to~$\vec{1}$, but we
cannot choose this eigenvector because it is forbidden by the condition
$\vec{1}^T\vec{x}=0$, which requires that the solution vector~$\vec{x}$ be
orthogonal to~$\vec{1}$.  (This is equivalent to saying that we're not
allowed to put all vertices in the same one group, which would certainly
ensure a small cut size, but wouldn't give a bisection of the graph.)  Our
next best choice is to choose $\vec{x}$ proportional to the eigenvector for
the second-lowest eigenvalue, the so-called \defn{Fiedler vector}.

\begin{figure}
\begin{center}
\includegraphics[width=7cm,clip=true]{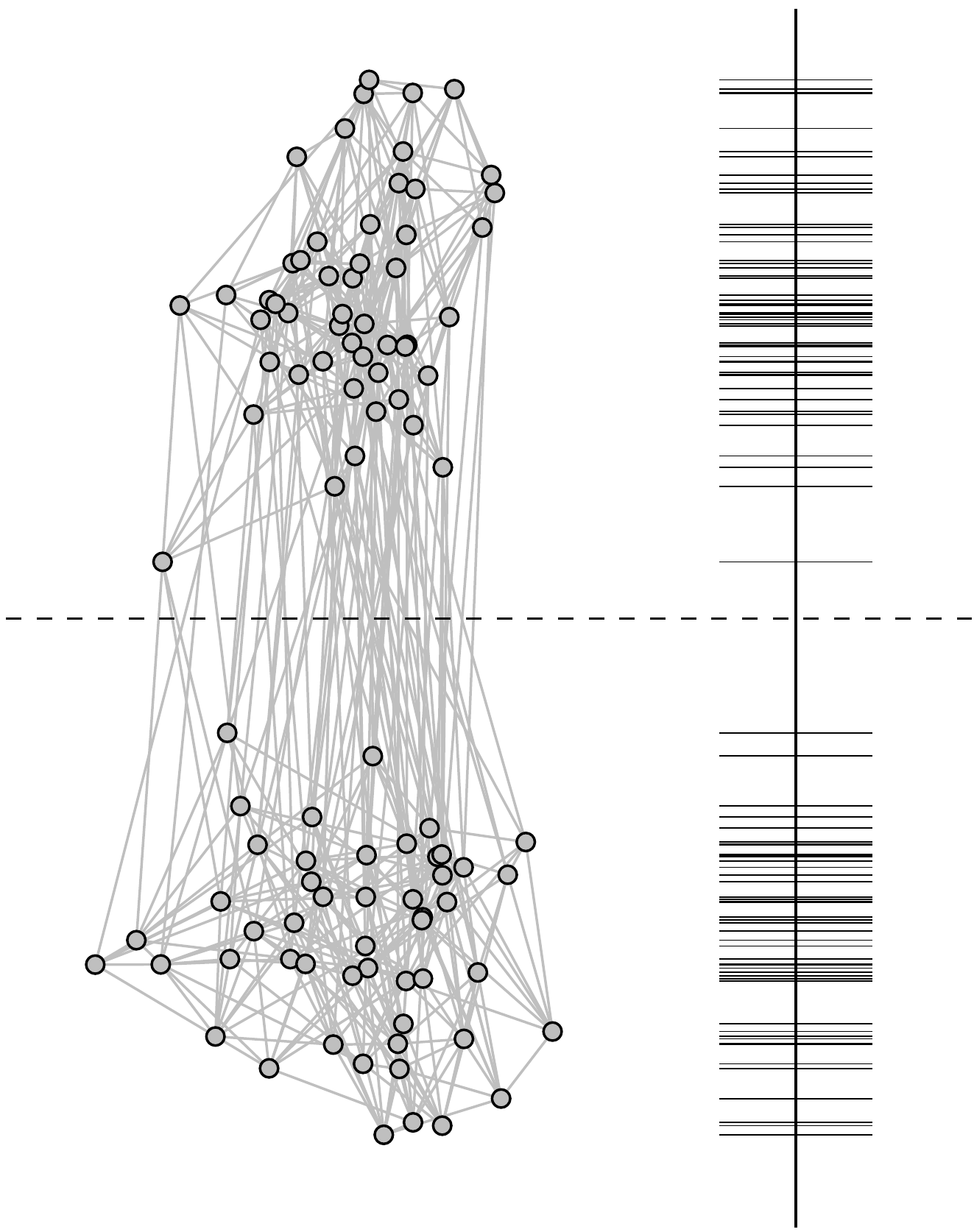}
\end{center}
\caption{Left: a small, computer-generated graph with two equally sized
  groups of vertices.  Right: the elements of the Fiedler vector---the
  second eigenvector of the graph Laplacian---plotted on an arbitrary
  scale.  A division of the vertices into two groups according to the signs
  of these elements (the dashed line indicates zero) recovers the groups.}
\label{fig:twoway}
\end{figure}

This solves the relaxed optimization problem exactly.  The final step in
the process is to ``unrelax'' back to the original variables~$s_i=\pm1$,
which we do by rounding the $x_i$ to the nearest value~$\pm1$, which means
that positive values of~$x_i$ are rounded to~$+1$ and negative values
to~$-1$.  Thus our final algorithm is a straightforward one: we calculate
the eigenvector of the graph Laplacian corresponding to the second-lowest
eigenvalue and then divide the vertices into two groups according to the
signs of the elements of this vector.  Although the solution of the relaxed
optimization is exact, the unrelaxation process is only an
approximation---there is no guarantee that rounding to $\pm1$ gives the
correct optimum for the unrelaxed problem---and hence the overall algorithm
only gives an approximate solution to the partitioning problem.  In
practice, however, it appears to work well.  Figure~\ref{fig:twoway} shows
an example.

As we have described it, the algorithm above also does not guarantee that
the two final groups of vertices will be of equal size.  The relaxed
optimization guarantees that $\sum_i x_i = 0$ because of
Eq.~\eqref{eq:sumx}, but we are not guaranteed that $\sum_i s_i = 0$ after
the rounding procedure.  Normally $\sum_i s_i$ will be close to zero, and
hence the groups will be of nearly equal sizes, but there may be some
imbalance.  In typical usage, however, this is not a problem.  In many
applications one is willing to put up with a small imbalance anyway, but if
one is not then a post-processing step can be performed that moves a small
number of vertices between groups in order to restore balance.

\section{Generalization to more than two groups}
\label{sec:kway}
Our primary purpose in this paper is to give a generalization of the
derivation of the previous section to spectral partitioning into more than
two groups.  The method we present allows the groups to be of any sizes we
choose---they need not be of equal size as in our two-way example.  As we
will see, the algorithm we derive differs in significant ways from previous
multiway partitioning algorithms.

\subsection{Cut size for multiway partitioning}
To generalize the spectral bisection algorithm to the case of more than two
groups we need first to find an appropriate generalization of the
quantities $s_i$ used in Section~\ref{sec:bisection} to denote membership
of the different communities.  For a partitioning into $k$ groups, we
propose using $(k-1)$-dimensional vectors to denote the groups, vectors
that in the simplest case point to the $k$ vertices of a
$(k-1)$-dimensional regular simplex.

\begin{figure}
\begin{center}
\includegraphics[width=\columnwidth]{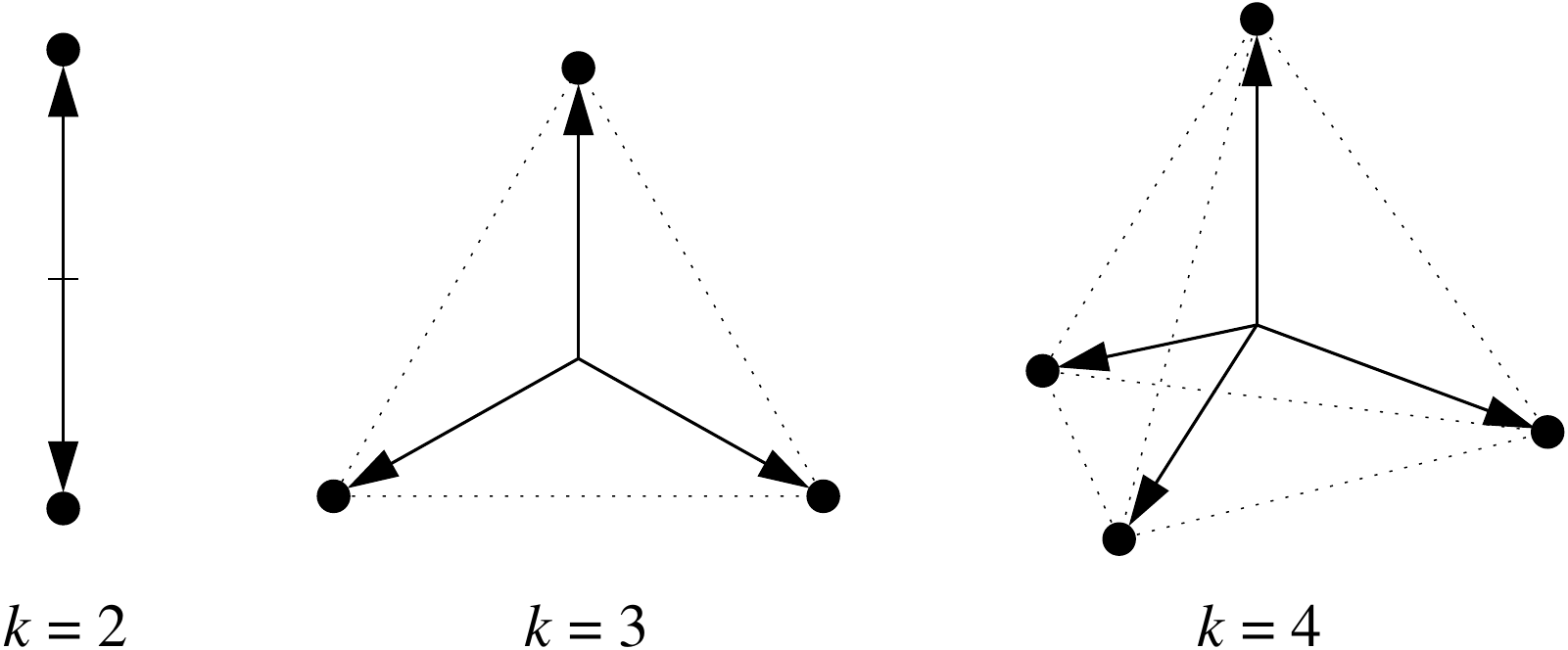}
\end{center}
\caption{As group labels we use vectors~$\vec{w}_r$ pointing from the
  center to the corners of a regular $(k-1)$-dimensional simplex.  For
  $k=2$ the simplex consists of just two points on a line, consistent with
  the indices $s_i=\pm1$ used in Section~\ref{sec:bisection}.  For $k=3$
  the simplex is an equilateral triangle; for $k=4$ it is a regular
  tetrahedron.  In higher dimensions it takes the form of the appropriate
  generalization of a tetrahedron to four or more dimensions, which would
  be difficult to draw on this two-dimensional page.}
\label{fig:simplices}
\end{figure}

Let us denote by $\vec{w}_r$ with $r=1\ldots k$ a set vectors pointing to
the vertices of a regular $(k-1)$-dimensional simplex centered on the
origin.  For $k=3$, for example, the three vectors would point to the
corners of an equilateral triangle; for $k=4$ the vectors would point to
the corners of a regular tetrahedron, and so forth---see
Fig.~\ref{fig:simplices}.  Such simplex vectors are not orthogonal.  Rather
they satisfy a relation of the form
\begin{equation}
\vec{w}_r^T\vec{w}_s = \delta_{rs} - {1\over k}.
\label{eq:vortho1}
\end{equation}
(Note that the individual vectors are normalized so that
$\vec{w}_r^T\vec{w}_r^{\vphantom{T}} = 1-1/k$.  One could normalize them to
have unit length, but subsequent formulas work out less neatly that way.)

We will use these simplex vectors as labels for the $k$ groups in our
partitioning problem, assigning one vector to represent each of the groups.
All assignments are equivalent and any choice will work equally well.  We
label each vertex~$i$ with a vector variable~$\vec{s}_i$ equal to the
simplex vector~$\vec{w}_r$ for the group~$r$ that it belongs to.  Then
Eq.~\eqref{eq:vortho1} implies that
\begin{equation}
\vec{s}_i^T\vec{s}_j + {1\over k}
  = \biggl\lbrace\begin{array}{ll}
      1 & \quad\mbox{if $i$ and $j$ are in the same group}, \\
      0 & \quad\mbox{otherwise,}
    \end{array}
\end{equation}
which is the equivalent of Eq.~\eqref{eq:sdelta}, and the derivation of the
cut size follows through as before, giving
\begin{equation}
R = \half \sum_{ij} (d_i \delta_{ij} - A_{ij} ) \vec{s}_i^T\vec{s}_j
  = \half \sum_{ij} L_{ij} \vec{s}_i^T\vec{s}_j,
\end{equation}
where $L_{ij}$ is once again an element of the graph Laplacian
matrix~$\mat{L}$.  Alternatively, we can introduce an $n\times(k-1)$
indicator matrix~$\tilde{\mat{S}}$ whose $i$th row is equal to~$\vec{s}_i$
and write the cut size in matrix notation as
\begin{equation}
R = \half \tr(\tilde{\mat{S}}^T\mat{L}\tilde{\mat{S}}).
\label{eq:cutsize}
\end{equation}

\begin{figure*}
\begin{center}
\includegraphics[width=\textwidth]{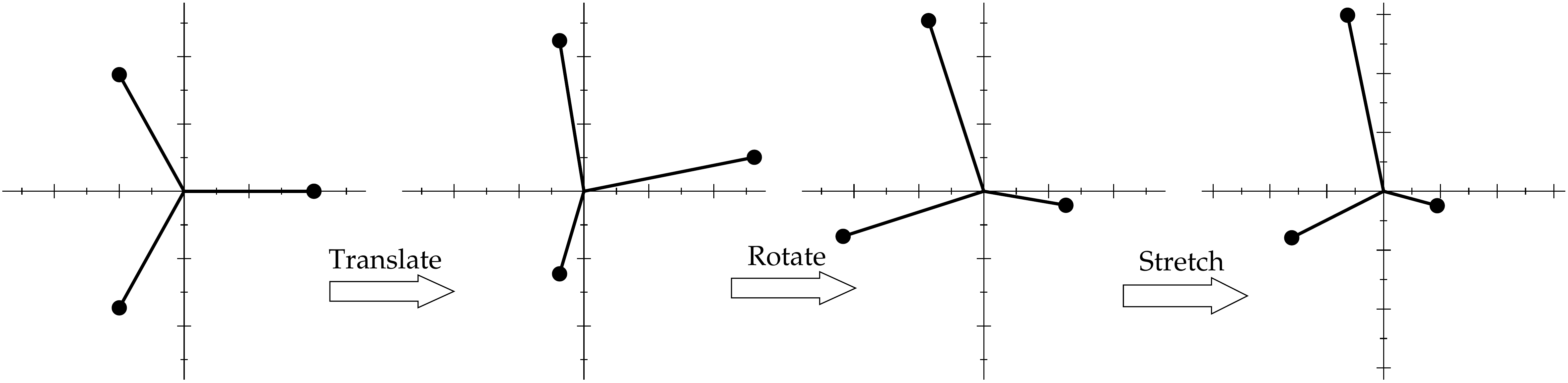}
\end{center}
\caption{As labels for the groups we use vectors derived from the regular
  simplex vectors of Fig.~\ref{fig:simplices} by a uniform translation,
  followed by a rotation and/or reflection, followed in turn by stretching
  or shrinking along each axis independently.}
\label{fig:transform}
\end{figure*}

These simplex vectors are not, however, the only possible choice for the
vectors~$\vec{s}_i$.  In fact, we have a lot of latitude about our choice.
The vectors can be translated, rotated, reflected, stretched or shrunk in a
variety of ways and still give a simple expression for the cut size.  If
the groups into which our graph is to be partitioned are of equal size,
then plain simplex vectors as above are a good choice, but for unequal
groups it will be useful to consider group vectors of the more general form
$\mat{D}\mat{Q}(\vec{w}_r-\vec{t})$, where $\mat{D}$ is a
$(k-1)\times(k-1)$ diagonal matrix, $\mat{Q}$~is a $(k-1)\times(k-1)$
orthogonal matrix, and $\vec{t}$ is an arbitrary vector.  This choice takes
the original simplex vectors and does three things, as illustrated in
Fig.~\ref{fig:transform}: first it translates them an arbitrary distance
given by~$\vec{t}$, then it rotates and/or reflects them according to the
orthogonal transformation~$\mat{Q}$, and finally it shrinks (or stretches)
them independently along each axis by factors given by the diagonal
elements of~$\mat{D}$.  The result is that the matrix~$\tilde{\mat{S}}$
describing the division of the network into groups is transformed into a
new matrix~$\mat{S}$:
\begin{equation}
\mat{S} = (\tilde{\mat{S}} - \vec{1}\vec{t}^T)\mat{Q}^T\mat{D}.
\label{eq:defss}
\end{equation}
Inverting this transformation, we get $\tilde{\mat{S}} =
\mat{S}\mat{D}^{-1}\mat{Q} + \vec{1}\vec{t}^T$ and substituting this
expression into Eq.~\eqref{eq:cutsize}, the cutsize can be written in terms
of the new matrix as
\begin{align}
R &= \half\tr\bigl[ (\mat{Q}^T\mat{D}^{-1}\mat{S}^T\!+\vec{t}\vec{1}^T) \mat{L}
                   (\mat{S}\mat{D}^{-1}\mat{Q}+\vec{1}\vec{t}^T) \bigr]
     \nonumber\\
  &= \half\tr(\mat{Q}^T\mat{D}^{-1}\mat{S}^T \mat{L} \mat{S}\mat{D}^{-1}\mat{Q})
   = \half\tr(\mat{S}^T \mat{L} \mat{S}\mat{D}^{-2}),
\end{align}
where in the second equality we have made use of $\mat{L}\vec{1}=0$.

The freedom of choice of the vector~$\vec{t}$ and the matrices~$\mat{Q}$
and~$\mat{D}$ allows us to simplify our problem as follows.  First, we will
require that $\vec{S}^T\mat{1} = 0$.  Taking the transpose of
Eq.~\eqref{eq:defss} and multiplying by~$\vec{1}$, this implies that
$(\tilde{\mat{S}}^T - \vec{t}\vec{1}^T)\vec{1} = 0$ and hence fixes the
value of~$\vec{t}$:
\begin{equation}
\vec{t} = {1\over n} \tilde{\mat{S}}^T\vec{1}
        = \sum_r {n_r\over n} \vec{w}_r,
\label{eq:choice1}
\end{equation}
where $n_r$ is the number of vertices in group~$r$.

The condition~$\vec{S}^T\mat{1} = 0$ is the equivalent of
Eq.~\eqref{eq:ssum} in the two-group case, and, just as~\eqref{eq:ssum}
does, it fixes the sizes of the groups, since the sum $\sum_r n_r
\vec{w}_r$ occupies a unique point in the space of the simplex vectors for
every choice of group sizes.

We would also like our matrix~$\mat{S}$ to satisfy a condition equivalent
to~$\vec{s}^T\vec{s}=n$ in the two-group case, which will take the form
$\mat{S}^T\vec{S} = \mat{I}$, where $\mat{I}$ is the identity matrix.  The
freedom to choose $\mat{Q}$ and $\mat{D}$ allows us to do this.  We note
that
\begin{equation}
\mat{S}^T\vec{S}
  = \mat{D}\mat{Q} (\tilde{\mat{S}}^T - \vec{t}\vec{1}^T)
     (\tilde{\mat{S}} - \vec{1}\vec{t}^T)\mat{Q}^T\mat{D},
\end{equation}
and that the central product in this expression is a $(k-1)\times(k-1)$
symmetric matrix that expands as
\begin{align}
(\tilde{\mat{S}}^T - \vec{t}\vec{1}^T) (\tilde{\mat{S}} - \vec{1}\vec{t}^T)
  &= \tilde{\mat{S}}^T\tilde{\mat{S}} - \tilde{\mat{S}}^T\vec{1}\vec{t}^T
     - \vec{t}\vec{1}^T\tilde{\mat{S}} + \vec{t}\vec{1}^T\vec{1}\vec{t}^T
     \nonumber\\
  &\hspace{-8em}{} =
     \tilde{\mat{S}}^T\tilde{\mat{S}} - n\vec{t}\vec{t}^T
   = \sum_r n_r \vec{w}_r^{\vphantom{T}}\vec{w}_r^T
     - \sum_{rs} {n_rn_s\over n}\vec{w}_r^{\vphantom{T}}\vec{w}_s^T,
\end{align}
where we have used~\eqref{eq:choice1}.

We perform an eigenvector decomposition of this matrix in the form
$\mat{U}\mat{\Delta}\mat{U}^T$, where where $\mat{U}$ is an orthogonal
matrix and $\mat{\Delta}$ is the diagonal matrix of eigenvalues, which are
all nonnegative since the original matrix is a perfect square.  Then we let
\begin{equation}
\mat{Q} = \mat{U}^T, \qquad \mat{D} = \mat{\Delta}^{-1/2},
\label{eq:choice2}
\end{equation}
and we have
\begin{align}
\mat{S}^T\vec{S}
  &= \mat{D}\mat{Q} (\tilde{\mat{S}}^T - \vec{t}\vec{1}^T)
     (\tilde{\mat{S}} - \vec{1}\vec{t}^T)\mat{Q}^T\mat{D} \nonumber\\
  &= \mat{\Delta}^{-1/2}\mat{U}^T \mat{U}\mat{\Delta}\mat{U}^T
     \mat{U}\mat{\Delta}^{-1/2} \nonumber\\
  &= \mat{\Delta}^{-1/2}\mat{\Delta}\mat{\Delta}^{-1/2}
   = \mat{I},
\end{align}
as required.

To summarize, we take simplex vectors centered on the origin and transform
them according to
\begin{equation}
\vec{w}_r \to \mat{D}\mat{Q}(\vec{w}_r-\vec{t}),
\label{eq:vectors}
\end{equation}
where $\vec{t}$, $\mat{Q}$, and~$\mat{D}$ are chosen according to
Eqs.~\eqref{eq:choice1} and~\eqref{eq:choice2}.  We use the transformed
vectors to form the rows of the matrix~$\mat{S}$, then the cut size for the
partition of the graph indicated by~$\mat{S}$ is given by
\begin{equation}
R = \half \tr(\mat{S}^T\mat{L}\mat{S}\mat{D}^{-2}),
\label{eq:cutsize2}
\end{equation}
while $\mat{S}$ obeys
\begin{equation}
\vec{S}^T\mat{1} = 0,
\label{eq:ones}
\end{equation}
and
\begin{equation}
\mat{S}^T\mat{S} = \mat{I}.
\label{eq:sortho}
\end{equation}

\subsection{Minimization of the cut size}
\label{sec:algorithm}
\begin{figure}
\begin{center}
\includegraphics[width=7cm]{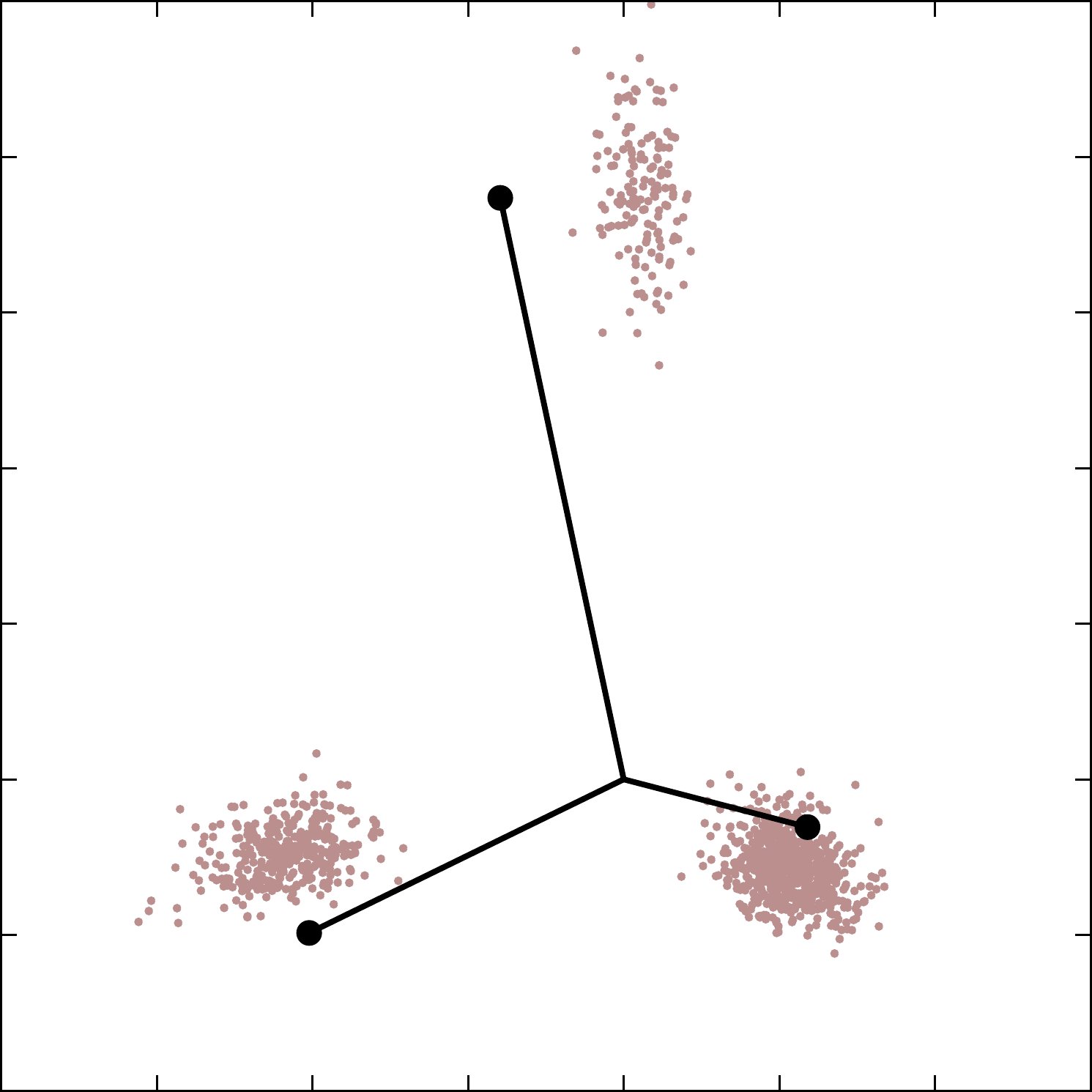}
\end{center}
\caption{The eigenvectors of the Laplacian define $n$ points in a
  $(k-1)$-dimensional space and we round each point to the nearest simplex
  vector to get an approximate solution to the partitioning problem.  The
  particular graph used here was, for the purposes of illustration,
  deliberately created with three groups in it of varying sizes, using a
  simple planted partition model~\cite{CK01} in which edges are placed
  between vertices independently at random, but with higher probability
  between vertices in the same group than between those in different
  groups.}
\label{fig:rounding}
\end{figure}

The remaining steps in the derivation are now straightforward, following
lines closely analogous to those for the two-group case.  Our goal is to
minimize the cut size~\eqref{eq:cutsize2} subject to the condition that the
group sizes take the desired value, which is equivalent to the
constraint~\eqref{eq:ones}.  Once again this is a hard computational
problem, but, by analogy with the two-group case, we can render it
tractable by relaxing the requirement that each row of~$\mat{S}$ be equal
to one of the discrete vectors~\eqref{eq:vectors}, solving this relaxed
problem exactly, then rounding to the nearest vector again to get an
approximate solution to the original unrelaxed problem.  The process is
illustrated in Fig.~\ref{fig:rounding}.

We replace~$\mat{S}$ with a matrix~$\mat{X}$ of continuous-valued elements
to give a relaxed cut size
\begin{equation}
R_x = \half \tr(\mat{X}^T\mat{L}\mat{X}\mat{D}^{-2}),
\label{eq:rx}
\end{equation}
where the elements of~$\mat{X}$ will be allowed to take any real values
subject to the constraint
\begin{equation}
\vec{X}^T\mat{1} = 0,
\label{eq:onex}
\end{equation}
equivalent to Eq.~\eqref{eq:ones}.  Again, however, we also need an
additional constraint, equivalent to Eq.~\eqref{eq:sumxsq}, to prevent all
elements of~$\mat{X}$ from becoming zero (which would certainly
minimize~$R_x$, but would not give a useful partition of the graph), and
the natural choice is the generalization of~\eqref{eq:sortho}:
\begin{equation}
\mat{X}^T\mat{X} = \mat{I}.
\label{eq:xortho}
\end{equation}
As in the two-group case, choices~of $\mat{X}$ satisfying this condition
necessarily include as a subset the original group vectors that
satisfy~\eqref{eq:sortho}, but also include many other choices as well.
Between them, the two conditions \eqref{eq:onex} and~\eqref{eq:xortho}
imply that the columns of~$\mat{X}$ should be orthogonal to one another,
orthogonal to the vector~$\vec{1}$, and normalized to have unit length.  As
we now show, the correct choice that satisfies all of these conditions and
minimizes~$R_x$ is to make the columns proportional to the eigenvectors of
the graph Laplacian corresponding to the second- to $k$th-lowest
eigenvalues.

The relaxed cut size~\eqref{eq:rx} can be minimized, as before, by
differentiating, applying the conditions \eqref{eq:onex}
and~\eqref{eq:xortho} with Lagrange multipliers:
\begin{align}
& {\partial\over\partial X_{kl}}
  \biggl[ \sum_{ijm} L_{ij} X_{im} X_{jm} D^{-2}_{mm} \nonumber\\
  &\hspace{5em} {} - \sum_{imn} \lambda_{mn} X_{im} X_{in}
  - \sum_{im} \mu_m X_{im} \biggr] = 0,
\end{align}
so that $2\sum_j L_{kj} X_{jl} D^{-2}_{ll} - 2\sum_m X_{km} \lambda_{ml} -
\mu_l = 0$, or in matrix notation $\mat{L}\mat{X}\mat{D}^{-2} =
\mat{X}\mat{\Lambda} + \half \vec{1}\bm{\mu}^T$, where $\mat{\Lambda}$ is a
$(k-1)\times(k-1)$ symmetric matrix of Lagrange multipliers and $\bm{\mu}$
is a $(k-1)$-dimensional vector.  As before, we can multiply on the left
by~$\vec{1}^T$ to show that $\bm{\mu}=0$, and hence we find that
\begin{equation}
\mat{L}\mat{X} = \mat{X}\mat{\Lambda}\mat{D}^2.
\label{eq:genev}
\end{equation}

Now, making use of the fact that $\mat{X}^T\mat{X}=\mat{I}$
(Eq.~\eqref{eq:xortho}), we have
\begin{equation}
\mat{\Lambda}\mat{D}^2 = \mat{X}^T\mat{X}\mat{\Lambda}\mat{D}^2
  = \mat{X}^T\mat{L}\mat{X} = \mat{D}^2\mat{\Lambda}\mat{X}^T\mat{X}
  =  \mat{D}^2\mat{\Lambda}.
\end{equation}
In other words, $\mat{D}^2$~and $\mat{\Lambda}$ commute.  Since $\mat{D}$
is diagonal this impiles that $\mat{\Lambda}$ is also diagonal, in which
case Eq.~\eqref{eq:genev} implies that each column of $\mat{X}$ is an
eigenvector of the graph Laplacian with the diagonal elements
of~$\mat{\Lambda}\mat{D}^2$ being the eigenvalues.  In other words the
eigenvalues are
\begin{equation}
\lambda_i = \Lambda_{ii}^{\vphantom{2}}D_{ii}^2.
\end{equation}
The conditions \eqref{eq:onex} and~\eqref{eq:xortho} tell us that the
eigenvectors must be distinct (because they are orthogonal to each other),
normalized to unity, and orthogonal to the vector~$\vec{1}$.

This still leaves us considerable latitude about which eigenvectors we use.
We can resolve the uncertainty by considering the cut size~$R_x$,
Eq.~\eqref{eq:rx}, which is given by
\begin{align}
R_x &= \half \tr(\mat{X}^T\mat{L}\mat{X}\mat{D}^{-2})
     = \half \tr(\mat{X}^T\mat{X}\mat{\Lambda})
     = \half \tr \mat{\Lambda} \nonumber\\
    &= \half \sum_i \Lambda_{ii} = \half \sum_i {\lambda_i\over D_{ii}^2}.
\label{eq:rxsoln}
\end{align}
Our goal is to minimize this quantity and, since both~$D_{ii}^2$ and the
eigenvalues of the Laplacian~$\lambda_i$ are nonnegative, the minimum is
achieved by choosing the smallest allowed eigenvalues of the Laplacian.  We
are forbidden by Eq.~\eqref{eq:onex} from choosing the lowest (zero)
eigenvalue, because its eigenvector is the vector~$\vec{1}$, so our best
allowed choice is to choose the columns of~$\mat{X}$ to be the eigenvectors
corresponding to the second- to $k$th-lowest eigenvalues of the Laplacian.
Which column is which depends on the values of the~$D_{ii}$.  The minimum
of~$R_x$ is achieved by pairing the largest~$\lambda_i$ with the
largest~$D_{ii}$, the second largest~$\lambda_i$ with the second
largest~$D_{ii}$, and so on.

This now specifies the value of the matrix~$\mat{X}$ completely and hence
consitutes a complete solution of the relaxed minimization problem.  The
correct choice of $\mat{X}$ is one in which the $k-1$ columns of the matrix
are equal to the normalized eigenvectors corresponding to the second- to
$k$th-lowest eigenvalues of the graph Laplacian, with the columns arranged
so that their eigenvalues increase in the same order as the diagonal
elements of the matrix~$\mat{D}$.

The only remaining step in the algorithm is to reverse the relaxation,
which means rounding the rows of the matrix~$\mat{X}$ to the nearest of the
group vectors---see Fig.~\ref{fig:rounding}.  As in the two-group case,
this introduces an approximation.  Although our solution of the relaxed
problem is exact, when we round to the nearest group vector there is no
guarantee that the result will be a true minimum of the unrelaxed problem.
Furthermore, as in the two-group case, we are not guaranteed that the
groups found using this method will be of exactly the required sizes~$n_r$.
The relaxed optimization must satisfy Eq.~\eqref{eq:onex}, but the
corresponding condition, Eq.~\eqref{eq:ones}, for the unrelaxed division of
the graph is normally only satisfied approximately and hence the groups
will only be approximately the correct size.  As in the two-group case,
however, this is typically not a problem.  Often we are content with an
approximate solution to the problem, but if not the groups can be balanced
using a post-processing step.  For example, the rounding of the relaxed
solution to the group vectors that preserves precisely the desired group
sizes can be calculated exactly in polynomial time using the so-called
Hungarian algorithm~\cite{PS98}, or approximately using a variety of vertex
moving heuristics.

\subsection{Practical considerations}
The method described in the previous section in principle constitutes a
complete algorithm for the approximate spectral solution of the minimum-cut
partitioning problem.  In practice, however, there are some additional
issues that arise in implementation.

First, note that the sign of the eigenvectors of the Laplacian is
arbitrary, and hence our matrix~$\mat{X}$ is only specified up to a change
of sign of any column, meaning there are $2^{k-1}$ choices of the matrix
that give equally good solutions to the relaxed optimization of the cut
size.  These $2^{k-1}$ solutions are reflections of one another in the axes
of the space occupied by the group vectors, and in practice the quality of
the solutions to the unrelaxed problem obtained by rounding each of these
reflections to the nearest group vector varies somewhat.  If we want the
best possible solution we need to look through all $2^{k-1}$ possibilities
to find which one is the best, and this could take a long time if $k$ is
large.

A second and more serious issue arises when the group sizes are equal to
one another, or nearly equal.  When the group sizes are equal the
conditions $\mat{S}^T\vec{1}=0$ and $\mat{S}^T\mat{S}=\mat{I}$ are
satisfied by the original, symmetric simplex vectors of
Fig.~\ref{fig:simplices} in any orientation.  This means that the group
vectors are not fully specified in this case---their orientation is
arbitrary.  When rounding the rows of the matrix~$\mat{X}$ to the nearest
simplex vector, therefore, an additional rotation may be required to find
the best solution.

\begin{figure}
\begin{center}
\includegraphics[width=7cm]{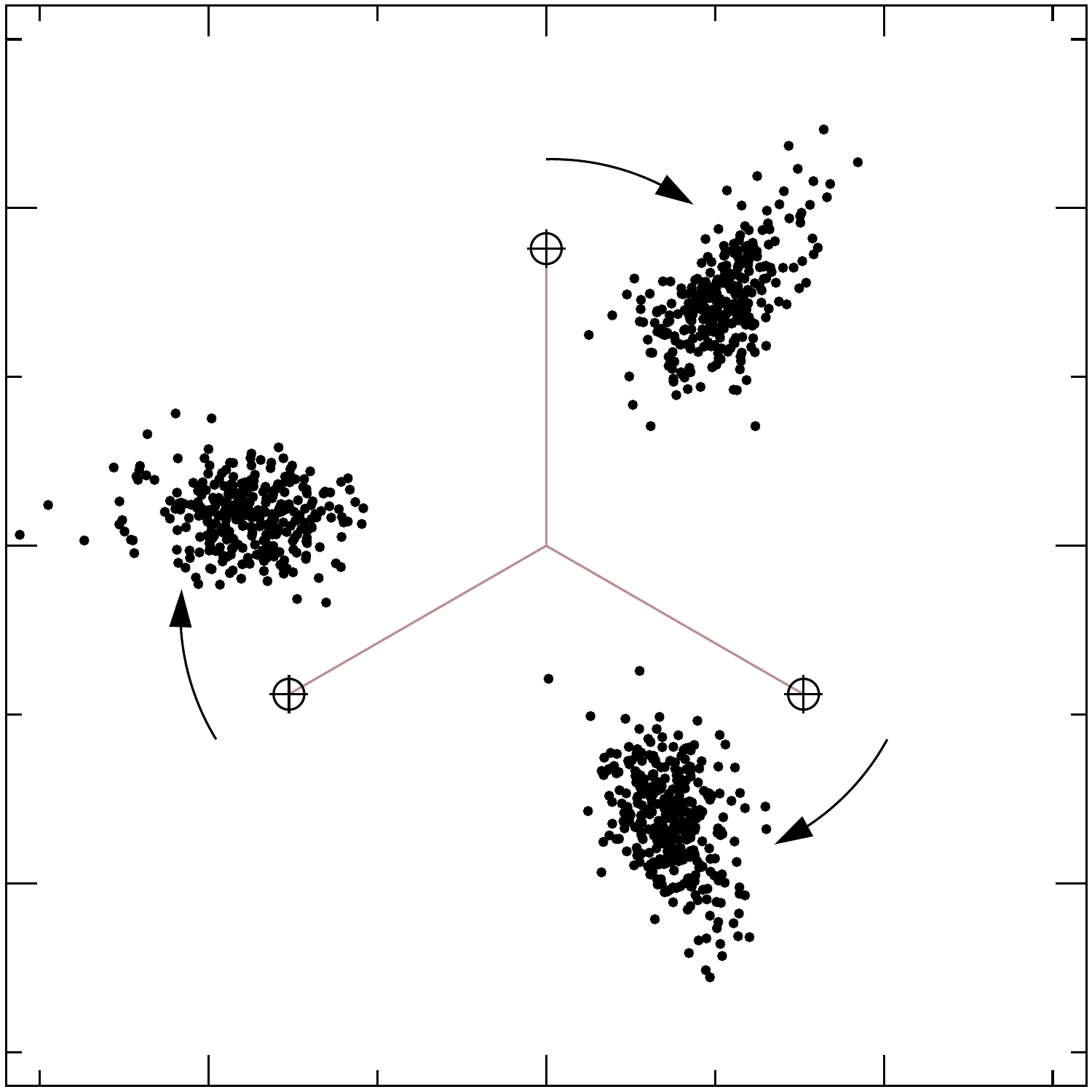}
\end{center}
\caption{The points in this plot represent the elements of the second and
  third eigenvectors of the Laplacian for a small graph of about a thousand
  vertices.  The graph used was, like that of Fig.~\ref{fig:rounding},
  created using a planted partition model, but with equally sized groups in
  this case.  The resulting points fall, roughly speaking, at the corners
  of a two-dimensional regular simplex, i.e.,~an equilateral triangle.  To
  determine the division of the graph into groups, we need to round these
  points to the nearest simplex vector, but the simplex must first be
  rotated to match the orientation of the points.}
\label{fig:rotate}
\end{figure}

The situation is depicted in Fig.~\ref{fig:rotate} for the case $k=3$.  The
rows~$\vec{x}_i$ are two-dimensional vectors in this case and form a
scatter of points in the plane of the plot as shown.  The points do indeed
approximate reasonably well to the corners of a regular simplex (an
equilateral triangle in this case), so in principle we should be able to
round them off and get a good solution to the partitioning problem.  But we
don't know \textit{a priori} what the correct orientation of the simplex
is, and in this case our first guess, as shown in the figure, is off and a
rotation is required.  We can rotate either the points or the simplex, but
we recommend rotating the simplex because it requires less work.

Given an assignment of vertices to groups, we can write down the
matrix~$\mat{S}$ of (unrotated) simplex vectors.  If we rotate the vectors,
this matrix becomes~$\mat{S}\mat{R}$, where $\mat{R}$ is a
$(k-1)\times(k-1)$ orthogonal matrix.  The sum of the squares of the
Euclidean distances from each point to the corresponding simplex vector is
given by
\begin{align}
\sum_{ij} [\mat{S}\mat{R}-\mat{X}]_{ij}^2
  &= \tr \bigl[ (\mat{R}^T\mat{S}^T-\mat{X}^T)(\mat{S}\mat{R}-\mat{X}) \bigr]
     \nonumber\\
  &= \tr \mat{S}^T\mat{S} - 2\tr \mat{R}^T\mat{S}^T\mat{X}
     + \tr \mat{X}^T\mat{X}.
\end{align}
The first and last terms in this expression are independent of~$\mat{R}$
and hence, for the purposes of choosing the rotation~$\mat{R}$ that
minimizes the whole expression, we need only minimize the middle term, or
equivalently maximize $\tr \mat{R}^T\mat{S}^T\mat{X}$.  The maximization of
this quantity over orthogonal matrices~$\mat{R}$ is a standard problem in
so-called Procrustes analysis~\cite{GD04}.  It can be solved by performing
a singular value decomposition of the matrix $\mat{S}^T\mat{X}$:
\begin{equation}
\mat{S}^T\mat{X} = \mat{U}\mat{\Sigma}\mat{V}^T,
\label{eq:svd}
\end{equation}
where $\mat{\Sigma}$ is the diagonal matrix of singular values and
$\mat{U}$ and $\mat{V}$ are orthogonal matrices.  Then
\begin{align}
\tr(\mat{R}^T\mat{S}^T\mat{X}) &= \tr(\mat{R}^T\mat{U}\mat{\Sigma}\mat{V}^T)
  = \tr(\mat{V}^T\mat{R}^T\mat{U}\mat{\Sigma}) \nonumber\\
  &\le \tr\mat{\Sigma},
\end{align}
the inequality following because $\mat{V}^T\mat{R}^T\mat{U}$, being a
product of orthogonal matrices, is also itself orthogonal, and all elements
of an orthogonal matrix are less than or equal to~1.  It is now trivial to
see that the exact equality---which is, by definition, the maximum of
$\tr(\mat{R}^T\mat{S}^T\mat{X})$ with respect to~$\mat{R}$---is achieved
when $\mat{R}^T = \mat{V}\mat{U}^T$ or equivalently when
\begin{equation}
\mat{R} = \mat{U}\mat{V}^T.
\label{eq:rotation}
\end{equation}
The product $\mat{U}\mat{V}^T$ is the orthogonal part of the \defn{polar
  decomposition} of $\mat{S}^T\mat{X}$.  Calculating it in practice
involves calculating first the singular value decomposition,
Eq.~\eqref{eq:svd}, and then discarding the diagonal matrix~$\mat{\Sigma}$.
Note that $\mat{S}^T\mat{X}$ is only a $(k-1)\times(k-1)$ matrix (not an
$n\times n$ matrix), and hence its singular value decomposition can be
calculated rapidly provided $k$ is small, in $\Ord(k^3)$ time.

These developments assume that we know the assignment of the vertices to
the groups.  In practice, however, we don't.  (If we did, we wouldn't need
to partition the graph in the first place.)  So in the algorithm we propose
we start with a random guess at the orientation of the simplex.  We round
the rows of $\mat{X}$ to the simplex vectors to determine group memberships
and then rotate the simplex vectors to fit the resulting groups according
to Eq.~\eqref{eq:rotation}.  We repeat this procedure until the groups no
longer change.  In a clear-cut case like that of Fig.~\ref{fig:rotate},
only one or two iterations would be needed for convergence, but in more
ambiguous cases we have found that as many as half a dozen or more
iterations may be necessary.

This algorithm works well for the case of exactly equal group sizes, while
the algorithm described in Section~\ref{sec:algorithm} works well for very
unbalanced groups, when the group vectors are specified completely, without
need for any rotation or Procrustes analysis.  A trickier scenario is when
the groups are almost but not exactly equal in size.  In such cases the
algorithm of Section~\ref{sec:algorithm} is correct in principle, but in
practice tends not to work very well---the particular orientation of the
group vectors picked by the algorithm may not agree well with the scatter
of points described by the rows of the matrix~$\mat{X}$.  In such cases, we
find that an additional Procrustes step to line up the points with the
group vectors usually helps.

But this raises the question of when the groups can be considered
sufficiently balanced in size that a possible rotation of the group vectors
may be needed.  Rather than try to answer this difficult question, we
recommend simply performing a Procrustes analysis and rotation for all
partitioning problems, whether one is needed or not.  In practice it
doesn't take long to do, and if it is not needed---if the points described
by the elements of~$\mat{X}$ are already well lined up with the group
vectors, as they are in Fig.~\ref{fig:rounding} for instance---then the
Procrustes analysis will simply do nothing.  It will leave the group
vectors unrotated (or rotate them only very slightly).

This approach has the added advantage of offering a solution to our other
problem as well, the problem of undetermined signs in the eigenvectors of
the Laplacian.  Since the orthogonal matrix~$\mat{R}$ in the Procrustes
analysis can embody a reflection as well as a rotation, the Procrustes
analysis will also determine which reflection gives the best fit of the
group vectors to the points, so we do not require an additional step to
deal with reflections.

Since the Procrustes analysis is an iterative method we do, in practice,
find that it can converge to the wrong minimum of the mean-square distance.
In the calculations presented in the remained of this paper, we run the
analysis several times with randomized starting conditions, taking the best
result over all runs, in order to mitigate this problem.

\subsection{Summary of the algorithm and running time}
Although the derivation of the previous sections is moderately lengthy, the
final algorithm is straightforward.  In summary the algorithm for
partitioning a given graph into $k$ groups of specified sizes is as
follows.
\begin{enumerate}
\item Generate a set of vectors~$k$ pointing to the vertices of a regular
  simplex centered at the origin and assign one vector as the label for
  each of the~$k$ groups.  Any orientation of the simplex can be used at
  this stage and any assignment of vectors to groups.
\item Define $\vec{t}$, $\mat{Q}$, and~$\mat{D}$ according to
  Eqs.~\eqref{eq:choice1} to~\eqref{eq:choice2}, then transform the simplex
  vectors according to
\begin{equation}
\vec{w}_r \to \mat{D}^{-1}\mat{Q}(\vec{w}_r-\vec{t}).
\end{equation}
\item Find the second- to $k$th-smallest eigenvalues of the graph
  Laplacian, and the corresponding normalized eigenvectors.  Pair the
  largest of these eigenvalues with the largest diagonal element
  of~$\mat{D}$, the second largest eigenvalue with the second largest
  diagonal element, and so forth.  Then form the matrix~$\mat{X}$, whose
  columns are the eigenvectors arranged in the same order as the diagonal
  elements of~$\mat{D}$ with which they are paired.
\item Rotate the group vectors~$\vec{w}_r$ into a random initial
  orientation.
\item Round each of the rows of $\mat{X}$ to the nearest group vector and
  construct the corresponding group matrix~$\mat{S}$ whose $i$th row is the
  group vector for the group that vertex~$i$ now belongs to.
\item Form the singular value decomposition $\mat{S}^T\mat{X} =
  \mat{U}\mat{\Sigma}\mat{V}^T$ and from it calculate the rotation matrix
  $\mat{R}=\mat{U}\mat{V}^T$.
\item Rotate the group vectors~$\vec{w}_r \to \vec{w}_r\mat{R}$.
\item Repeat from step~5 until group memberships no longer change.
\end{enumerate}

Most often we are interested in sparse graphs in which the number of edges
is proportional to the number of vertices, so that the mean degree of a
vertex tends to a constant as the graph becomes large.  In this situation
the eigenvectors of the Laplacian can be calculated using sparse iterative
methods such as the Lanczos algorithm.  The Lanczos algorithm takes
time~$\Ord(k^2n)$ per iteration, and although there are no formal results
for the number of iterations required for convergence, the number in
practice seems to be small.  The other steps of the algorithm all also take
time~$\Ord(k^2n)$ or less, and hence the algorithm has leading-order
worst-case running time~$\Ord(k^2 n)$ times the number of Lanczos
iterations, making it about as good as traditional approaches based on
$k$-means, and well suited for large graphs.  (Formal results for the
number of iterations $k$-means takes to converge are also not available, so
a precise comparison of the complexity of the two methods is not possible.)

\subsection{Weighted graphs and data clustering}
\label{sec:weighted}
The methods described in the previous sections can be extended in a
straightforward manner to weighted graphs---graphs with edges of varying
strength represented by varying elements in the adjacency matrix.  For such
graphs the goal of partitioning is to divide the vertices into groups such
that the sum of the weights of the edges running between groups is
minimized.  To achieve this we generalize the degree~$d_i$ of vertex~$i$ in
the obvious fashion $d_i = \sum_j A_{ij}$ and the elements of the Laplacian
accordingly $L_{ij} = d_i \delta_{ij} - A_{ij}$.  Then the cut size once
again satisfies Eq.~\eqref{eq:cutsize}, and the rest of the algorithm
follows as before.  We have not experimented extensively with applications
to weighted graphs, but in preliminary tests the results look promising.

One can also apply our methods to the problem of data clustering, the
grouping of points within a multidimensional data space into clusters of
similar values~\cite{VM03b,vonLuxburg07}.  One standard approach to this
problem makes use of an affinity matrix.  Suppose one has a set of $n$
points represented by vectors~$\vec{r}_i$ in a $d$-dimensional data space.
One then defines the affinity matrix~$\mat{A}$ to have elements
\begin{equation}
A_{ij} = \e^{-|\vec{r}_i-\vec{r}_j|^2/2\sigma^2},
\end{equation}
where $\sigma$ is a free parameter chosen by the user.  If $\sigma$ is
roughly of order the distance between intra-cluster points, then $A_{ij}$
will approximately take the form of the adjacency matrix of a weighted
graph in which vertices are connected by strong edges if the corresponding
data points are near neighbors in the data space.  (For values of $\sigma$
much larger or smaller than this clustering methods based on the affinity
matrix will not work well, so some care in choosing $\sigma$ is necessary
to get good results.  Automated methods have been proposed for choosing a
good value~\cite{NJW01}.)

Given the affinity matrix, we can now apply the method described above for
weighted graphs to this matrix and derive a clustering of the data points.
We will not pursue this idea further in the present paper, but in
preliminary experiments on standard benchmark data sets we have found that
the algorithm gives results comparable with, and in some cases better than,
other spectral clustering methods.

\section{Results}
\label{sec:results}
Our primary purpose in this paper is to provide a first-principles
derivation of a multiway spectral partitioning algorithm.  However, given
that the algorithm we have derived differs from standard algorithms, it is
also of interest to examine how well it performs in practice.  In this
section we give example applications of the algorithm to graphs from a
variety of sources.  Our tests do not amount to an exhaustive
characterization of performance, but they give a good idea of the basic
behavior of the algorithm.  Overall, we find that the algorithm has
performance comparable to that of other spectral algorithms based on
Laplacian eigenvectors, but there exist classes of graphs for which our
algorithm does measurably better.  In particular, the algorithm appears to
perform better than some competitors in cases where the partitioning task
is particularly difficult.

\begin{figure}
\begin{center}
\includegraphics[width=7.5cm,clip=true]{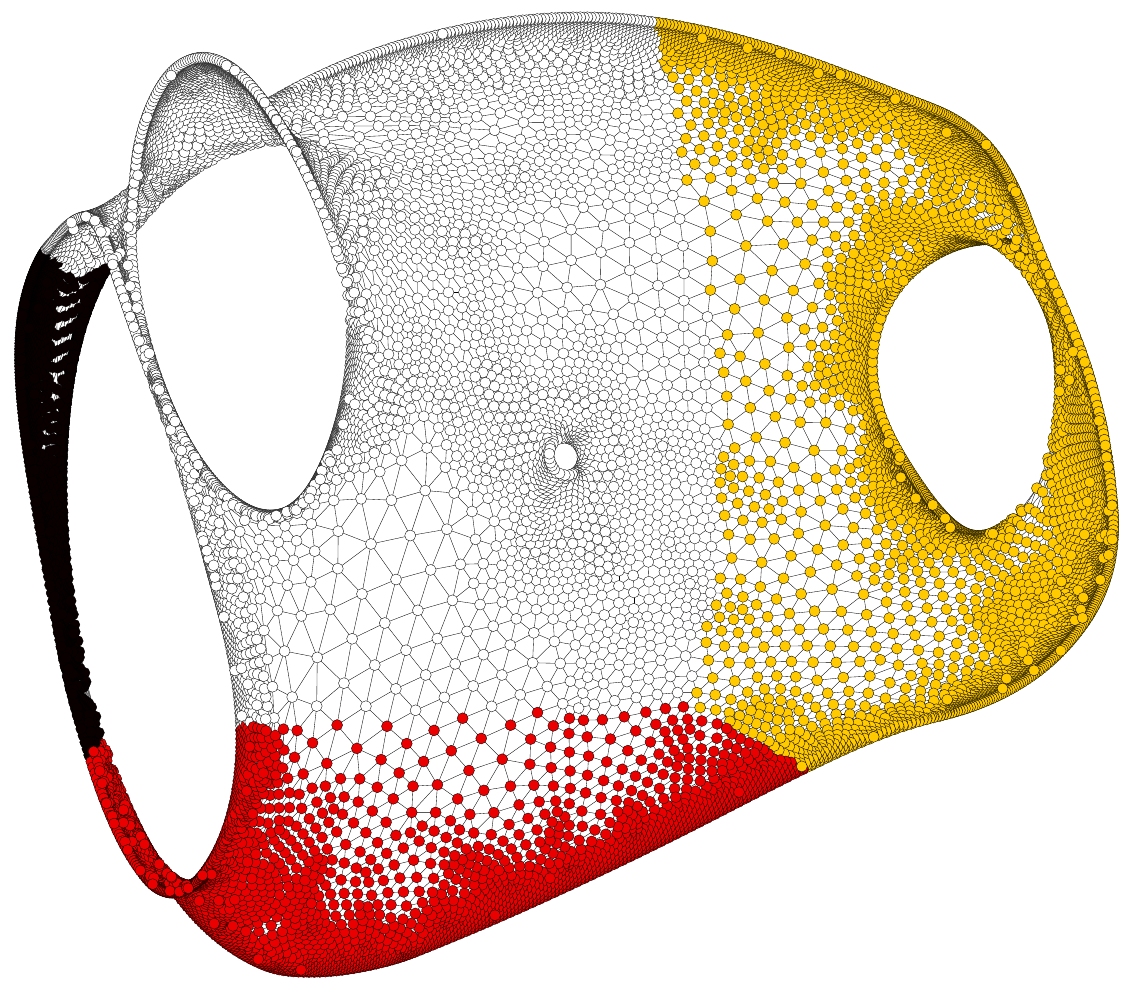}
\end{center}
\caption{Division of a structural engineering mesh network of $15\,606$
  vertices into four parts---represented by the four colors---using the
  algorithm described in this paper.  The sizes of the parts in this case
  were 1548, 2745, 4979, and 6334.  The complete graph has $45\,878$ edges;
  this division cuts just 351 of them, less than 1\%.  Graph data courtesy
  of the University of Florida Sparse Matrix Collection.}
\label{fig:mesh}
\end{figure}

As a first example, Fig.~\ref{fig:mesh} shows the result of applying our
algorithm to a graph from the University of Florida Sparse Matrix
Collection.  This graph is a two-dimensional mesh network drawn from a NASA
structural engineering computation, and is typical of finite-element meshes
used in such calculations (which are a primary application of partitioning
methods).  Figure~\ref{fig:mesh} shows a split of the graph into four parts
of widely varying sizes.  The split is closely similar to that found by
conventional spectral partitioning using $k$-means, indicating that our
algorithm has comparable performance on this application to standard
methods in current use.

\begin{figure}
\begin{center}
\includegraphics[width=8cm,clip=true]{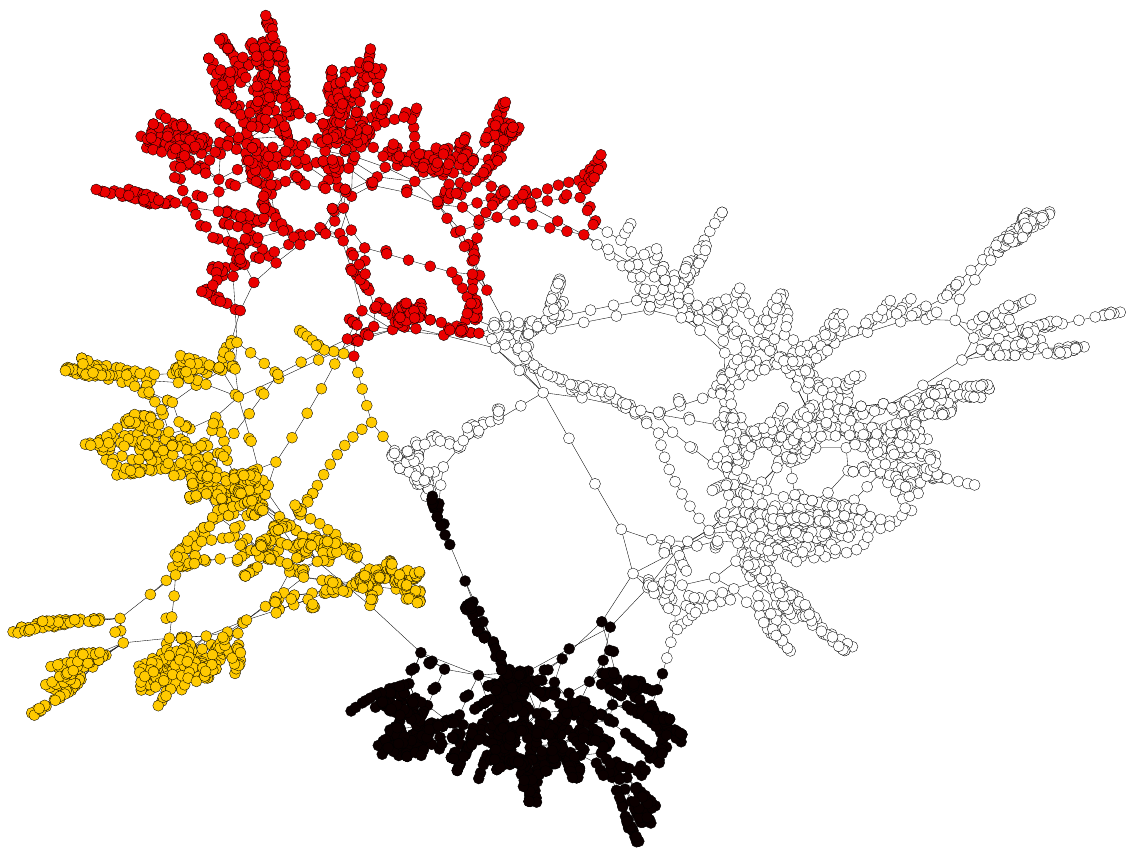}
\end{center}
\caption{Division into four parts of a 4941-vertex graph representing the
  Western States Power Grid of the United States.  The sizes of the parts
  were 898, 1066, 1240, and 1737.  The complete graph contains 6594 edges,
  of which 25 are cut in this division.  Graph data courtesy of Duncan
  Watts.}
\label{fig:grid}
\end{figure}

Figure~\ref{fig:grid} shows an application to a graph representing a power
grid, specifically the Western States Power Grid, which is the network of
high-voltage electricity transmission lines that serves the western part of
the United States~\cite{WS98}.  The figure shows the result of splitting
the graph into four parts and the split is an intuitively sensible one and
again comparable to that found using more traditional methods.

There are, however, some graphs for which our method gives results that are
significantly better than those given by previous methods, particularly
when the target group sizes are significantly unbalanced.  As a controlled
test of the performance of the algorithm we have applied it to artificial
graphs generated using a planted partition model~\cite{CK01} (also called a
stochastic block model in the statistical literature~\cite{SN97b}).  In
this model one creates graphs with known partitions by dividing a specified
number of vertices into groups and then placing edges within and between
those groups independently with given probabilities.  In our tests we
generated graphs of 3600 vertices with three groups.  Edges were placed
between vertices with two different probabilities, one for vertices in the
same group and one for vertices in different groups, chosen so that the
average degree of a vertex remained constant at~40.  We then varied the
fraction of edges placed within groups to test the performance of the
algorithm.

\begin{figure*}
\begin{center}
\includegraphics[width=16cm]{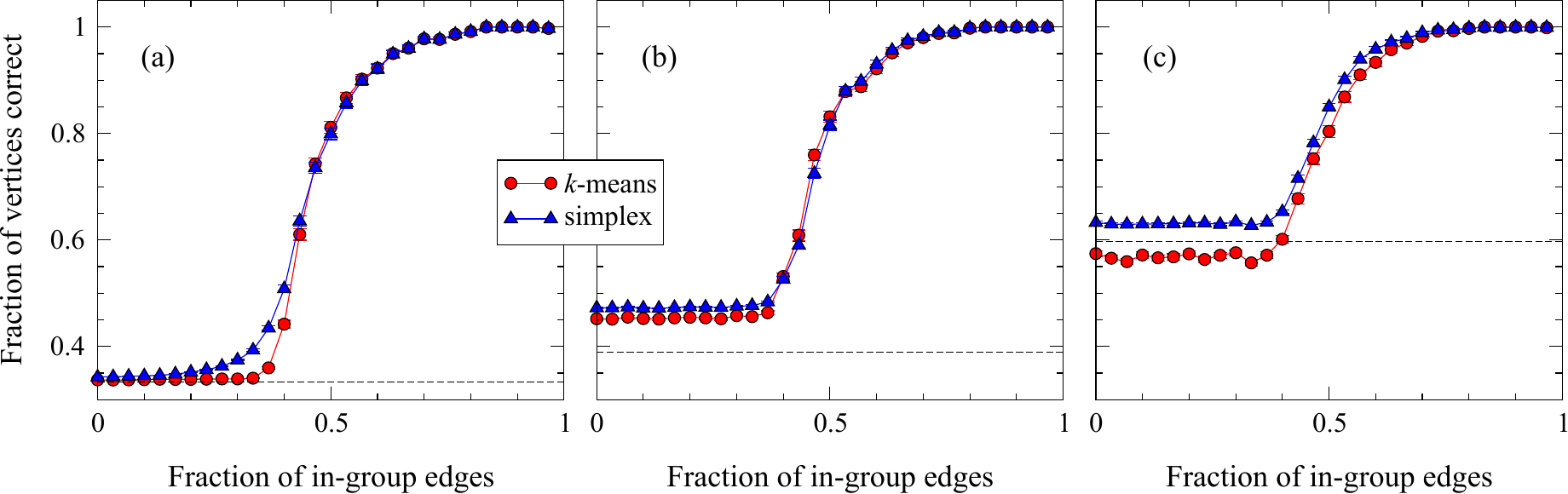}
\end{center}
\caption{Fraction of vertices classified into the correct groups by a
  standard spectral algorithm based on $k$-means (red circles) and by the
  algorithm described in this paper (blue triangles), when applied to
  graphs of 3600 vertices, artificially generated using a planted partition
  model with three groups.  In (a) the groups are of equal sizes.  In (b)
  the sizes are 1800, 1200, and~600.  In (c) they are 2400, 900, and~300.
  The dashed horizontal line in each frame represents the point at which
  the algorithms do no better than chance.  Each data point is an average
  over 500 networks and the calculation for each network is repeated with
  random initial conditions as described in the text; the results shown
  here are the best out of ten such repeats.}
\label{fig:results}
\end{figure*}

Figure~\ref{fig:results} shows the results of applying both our algorithm
and a standard $k$-means spectral algorithm to a large set of graphs
generated using this method.  We compare the divisions found by each
algorithm to the known correct divisions and calculate the fraction of
vertices classified into the correct groups as a function of the fraction
of in-group edges.  When the latter fraction is large the group structure
in the network should be clear and we expect any partitioning algorithm to
do a good job of finding the best cut.  As the fraction of in-group edges
is lowered, however, the task gets harder and the fraction of correct
vertices declines for both algorithms, eventually approaching the value
represented by the dashed horizontal lines in the figure, which is the
point at which the classification is no better than chance---we would
expect a random division of vertices to get about this many vertices right
just by luck.  (If group~$i$ occupies a fraction~$\nu_i$ of the network,
then a random division into groups of the given sizes will on average get
$\sum_i \nu_i^2$ vertices correct.)

The first panel in the figure shows results for groups of equal size and
for this case the performance of the two algorithms is similar.  Both do
little better than random for low values of the fraction of in-group edges.
The simplex algorithm of this paper performs slightly better in the hard
regime, but the difference is small.  When the group sizes are different,
however, our algorithm outperforms the $k$-means algorithm, as shown in the
second and third panels.  In the third panel in particular, where the group
sizes are strongly unbalanced, our algorithm performs substantially better
than $k$-means for all parameter values, but particularly in the hard
regime where the fraction of in-group edges is small.  In this regime the
$k$-means algorithm does no better than a random guess, but our
simplex-based algorithm does significantly better.

To be fair, we should also point out that there are some cases in which the
$k$-means algorithm outperforms the algorithm of this paper.  In
particular, we find that in tests using the planted partition model with
three groups of equal sizes, but where the between-group connections are
asymmetric and one pair of groups is more weakly connected than the other
two pairs, the $k$-means algorithm does better in certain parameter
regimes.  The explanation for this phenomenon appears to be that our
algorithm has difficulty finding the best orientation of the simplex to
perform the partitioning.  It is possible that one could achieve better
results using a different method for finding the orientation other than the
Procrustes method used here.  The $k$-means partitioning algorithm, which
does not use an orientation step, has no corresponding issues.

\section{Conclusions}
\label{sec:conclusions}
In this paper, we have derived a multiway spectral partitioning algorithm
from first principles as a relaxation approximation to a well-defined
minimum-cut problem.  This contrasts with more traditional presentations in
which an algorithm is proposed \textit{ex nihilo} and then proved after the
fact to give good results.  While both approaches have merit, ours offers
an alternative viewpoint that helps explain why spectral algorithms
work---because the spectral algorithm is, in a specific sense, an
approximation to the problem of minimizing the cut size over divisions of
the graph.

Our approach not only offers a new derivation, however; the end product,
the algorithm itself, is also different from previous algorithms, involving
a vector representation of the partition with the geometry of an irregular
simplex.  In practice, the algorithm appears to give results that are
comparable with those of previous algorithms and in some cases better.  The
algorithm is also efficient.  For graphs of $n$ vertices divided into $k$
groups, the running time is dictated by the calculation of the eigenvectors
of the graph Laplacian matrix, which for a sparse graph can be done using
the Lanczos method in time $\Ord(k^2n)$ times the number of Lanczos
iterations (which is typically small), so overall running time is roughly
linear in~$n$ for given~$k$.

The developments described here leave some questions unanswered.  In
particular, our method fixes the group sizes within the relaxed
approximation to the minimization problem, but in the true problem the
sizes are only fixed approximately.  A common variant of the minimum-cut
problem arises when the group sizes are not exactly equal but are allowed
to vary within certain limits.  Our method could be used to tackle this
problem as well, but one would need a scheme for preventing the size
variation from passing outside the allowed bounds.  These and related ideas
we leave for future work.

\begin{acknowledgments}
  The authors thank Raj Rao Nadakuditi for useful insights and suggestions.
  This work was funded in part by the National Science Foundation under
  grant DMS--1107796.
\end{acknowledgments}

\end{document}